\documentclass[sn-mathphys-num]{sn-jnl}

\usepackage{multirow}%
\usepackage{graphicx}%
\usepackage{amsmath,amssymb,amsfonts}%
\usepackage{amsthm}%
\usepackage{mathrsfs}%
\usepackage[title]{appendix}%
\usepackage{xcolor}%
\usepackage{textcomp}%
\usepackage{manyfoot}%
\usepackage{booktabs}%
\usepackage{algorithm}%
\usepackage{algorithmicx}%
\usepackage{algpseudocode}%
\usepackage{listings}%

\usepackage[T1]{fontenc}
\usepackage[utf8]{inputenc}

\usepackage{mathtools}

\usepackage{float}
\usepackage{etoolbox}
\usepackage{longtable}
\AtBeginEnvironment{longtable}{\tiny}

\newcommand{\N}{\mathbb{N}}
\newcommand{\Z}{\mathbb{Z}}
\newcommand{\BB}{{\mathcal B}}
\newcommand{\rr}{{\rho}}
\usepackage{array,multirow,makecell}
\setcellgapes{1pt}
\makegapedcells
\newcolumntype{R}[1]{>{\raggedleft\arraybackslash }b{#1}}
\newcolumntype{L}[1]{>{\raggedright\arraybackslash }b{#1}}
\newcolumntype{C}[1]{>{\centering\arraybackslash }b{#1}}

\begin{document}

\title[CA Controllability as SAT Problems]{Regional Controllability of Cellular Automata  as a SAT Problem}

\author*[1,2]{Franco Bagnoli}\email{franco.bagnoli@unifi.it}

\author[3]{Sara Dridi}\email{sara.dridi@univ-setif.dz}
\equalcont{These authors contributed equally to this work.}

\author[4]{Nazim Fatès}\email{nazim.fates@loria.fr}
\equalcont{These authors contributed equally to this work.}

\affil*[1]{Department Physics and Astronomy and CSDC, University of Florence, Via G. Sansone,1,  Sesto Fiorentino, 50019,  Italy}

\affil[2]{INFN, Sect. Florence}

\affil[3]{ Institute of Optics and Precision Mechanics,  University of Setif 1, Algeria}

\affil[4]{Université de Lorraine, CNRS, Inria, LORIA, Nancy, 54000,  France}

\abstract{
Controllability, one of the fundamental concepts in control theory, consists in guiding a system from an initial state to a desired one within a limited (and possibly minimum) time interval. When the objective is limited to a specific sub-region of the system’s domain, the concept is referred to as regional controllability.

We examine this notion in the context of Boolean one-dimensional cellular automata of finite length. Depending on the local evolution rule, we investigate whether it is possible to control the evolution of the system by imposing particular values on the boundary conditions. 
This approach is related to key dynamical properties of CA, specifically chain transitivity and chain mixing.
We show that the control problem can be formulated as a Boolean satisfiability (SAT) problem and can thus be addressed using SAT solvers.  
We also show how finding shortest paths in the configuration graph  allows to determine controllability properties. From our observations we can state that only peripherally-linear rules are fully controllable, while for other rules, the reachability ratio, that is, the fraction of controllable pairs of initial and final configurations, is vanishing when the system size grows. 
}


\maketitle

%make life simpler: @ => bold
\let\at=@
\catcode`\@=\active
\def@#1{\ifmmode\boldsymbol{#1}\else\at#1\fi}

%%%%%%%%%%%%%%%%%%%%%%%%%%%%%%%%%%%%%%%%%
\section{Introduction}\label{sec:intro}

Control theory is a branch of mathematics and engineering that deals with the behaviour of dynamical systems and how this behaviour can be modified with some influence that takes the form of a feedback. In other words, to control a system means to influence its behaviour so as to steer it to a desired state. Control theory is widely used in large number of fields ranging from aerospace, robotics, electrical engineering to social science.  

Many studies have explored the control of dynamical systems with continuous variables and continuous-time evolution addressing both finite- and infinite-dimensional cases. These systems are typically modelled and analysed using differential equations and partial differential equations~\cite{lions1991exact,curtain2012introduction}.

Controllability, introduced by Kalman in 1960, is one of the fundamental concepts in control theory.  It explores whether a system can be guided from any initial state to a desired state within a predefined time interval $[0,T]$. Since the, this concept has been extensively studied for finite-dimensional systems \cite{sontag2013mathematical} and infinite-dimensional systems described by partial differential equations (PDE) \cite{lions1968controle,lions1986controlabilite}.  

More recently, controllability has been studied in the context of cellular automata (CA -- acronym also used to define single cellular automaton), which are considered as potential alternatives to classical models based on partial differential equations, as they can effectively capture non-linear phenomena through simple local rules. 

Cellular automata are discrete dynamical systems regarded as  the simplest models of spatially extended systems which can offer an effective framework for describing complex phenomena.  They consist of three components: a grid of cells, each taking a state from a finite set, a neighbourhood (for each cell) and a local transition function. The CA paradigm has been successfully applied to a wide range of fields, including biology, chemistry, physics, and ecology, as evidenced by an extensive literature on the subject~\cite{ermentrout1993cellular,kier2005cellular,xiao2005using}. For a general overview, see the proceedings of the ACRI conference~\cite{ACRI}. 

The focus of this paper is on a specific aspect of controllability, known as regional controllability, where the goal is to achieve a desired objective only on a part of the whole domain by applying actions on its boundaries.   Regional controllability of cellular automata focuses on the ability to guide the state of a system toward desired configurations within a specific subregion of the entire domain.  
This concept was introduced by El Jai and Zerrik in (1993-1995) and well studied in  a wide range of works in the context of distributed parameter systems described by partial differential equations. It is particularly relevant for this type of systems to address situations where full domain control is either unachievable or unnecessary, while control within a specific region remains achievable. 

When switching from a continuous description (PDE) to a discrete one (coupled maps or CA), we have also to introduce the system size $n$ as a relevant parameter.

In the context of CA, different characterization results have been proposed to extend or substitute the widely used Kalman criterion and the regional controllability problem was analysed with various methods to establish new criteria for the case of cellular automata and discrete complex systems~\cite{dridi2019recent}.

Our research primarily focused on the regional controllability of Boolean CA, demonstrating its validity through the use of Markov chains and graph theory tools \cite{dridi2019graph,dridi2019markov,dridi2020boundary}; see Ref.~\cite{Bagnoli2025} for a general overview. 

The problem of regional controllability  has been addressed with an approach based on the Kalman condition~\cite{dridi2022kalman,el2021some}. 
For one-dimensional deterministic cellular automata, the problem has been explored using the concept of Boolean derivatives~\cite{bagnoli2016toward} and the probabilistic case has been investigated with a Markov-chain approach~\cite{bagnoli2018regional,bagnoli2019optimal}. In a recent work, a novel characterization of controllability and regional controllability based on symbolic dynamics was introduced~\cite{dridi2025new}.  This study establishes that the regional controllability for every $n$ (where $n$ is the size of the controlled region) and controllability of cellular automata are equivalent to the topological properties, namely those of chain transitivity and chain mixing.

In another recent work, a preimage algorithm was used to determine whether a desired configuration can be reached from an initial configuration, using a characterization tool known as the \textit{controllability tree}~\cite{Dridi2024regional}. This paper builds upon these findings. 

In this work, we deepen our exploration of the regional controllability of elementary cellular automata by formulating this problem as a satisfiability problem (SAT). 
We also provide new techniques such as finding the minimum path in the directed configuration graph.

The structure of the paper is as follows. Section~\ref{sec:2} provides an overview of the definitions of elementary cellular automata,  introduces the regional controllability problem and the connection between chain transitivity, chain mixing and regional controllability. Section~\ref{sec:3} formulates the problem of regional controllability for one-dimensional cellular automata as a SAT problem. Furthermore, this method can also be used to explore key dynamical properties of cellular automata, particularly chain transitivity and chain mixing. Section~\ref{sec:4} is devoted to the generation of preimages of a given configuration, given the controls, and in Section~\ref{sec:tree}, we present another technique for CA control, focusing on the process of finding the shortest path in the directed configuration graph, achieved by constructing trees of images and preimages corresponding to the starting and target configurations for all possible control inputs. Finally,  conclusions are drawn in Section \ref{sec:5}. 

%%%%%%%%%%%%%%%%%%%%%%%%%%%%%%%%%%%%%%%%%%%%%%%%%%%%%%%%%%%%%%%%%%%%
\section{Definitions of the problem and mathematical properties}\label{sec:2}

\subsection{Elementary Cellular Automata}

Elementary cellular automata (ECA -- acronym also used to define a single automaton) are discrete dynamical systems for which the state of each cell only takes two values and is determined at each time step by its own state and the state of left and right neighbours. The evolution of the states of all cells occurs in parallel. 

We first consider a one-dimensional infinite set of cells. The state of each cell $i \in \Z $ at time $t$ is given by a variable $x_i^{t}\in\{0,1\}$.  Mathematically, the evolution of the cells is defined in terms of a local function $f \{0,1\}^3 \rightarrow \{0,1\}: $
\[
          \forall i \in \Z, x^{t+1}_i =f(x^t_{i-1},x^t_{i},x^t_{i+1}).
\]

For a given time $ t$, we denote by $ @x^t = (x^{t}_i)_{i\in\Z} $ the sequence of all cell states, i.e., a {\em configuration}. In the rest of the paper we consider only finite configurations since we will focus on a particular set of cells, or a {\em region}, not taking into consideration what happens outside it. 

\begin{table}[h]
  \caption{\label{tab:150} The look-up table of Rules 150 and 22. $\mathcal{N}$ is the neighbourhood in the base-two representation, $\mathcal{N}^{(10)}$ is the same in then base-ten representation. The column $f^{(150)}$ is the output of Rule 150 and $D^{(150)}_\mathcal{N}$ the derivative of the Rule 150 in zero with respect to the ones in the neighbourhood in the base-two representation, same for $f^{(22)}$ and $D^{(22)}$ for Rule 22.}
	 %\begin{center}
    \begin{tabular}{c|c|c|c|c|c}
 $\mathcal{N}=(x_{-1} , x_{ 0} , x_{ 1} )$& $\mathcal{N}^{(10)}=4x_{-1}+2x_0+x_1$& $f^{(150)}(\mathcal{N})$ &$D^{(150)}_\mathcal{N}$& $f^{(22)}(\mathcal{N})$ &$D^{(22)}_\mathcal{N}$\\
 \hline
      $0,0,0$ & $0$& $0$ & $0$& $0$ & $0$\\
      $0,0,1$ & $1$& $1$& $1$& $1$ & $1$\\
      $0,1,0$ & $2$& $1$& $1$& $1$ & $1$\\
      $0,1,1$ & $3$& $0$& $0$& $0$ & $0$\\
      $1,0,0$ & $4$& $1$& $1$& $1$ & $1$\\
      $1,0,1$ & $5$& $0$& $0$& $0$ & $0$\\
      $1,1,0$ & $6$& $0$& $0$& $0$ & $0$\\
      $1,1,1$ & $7$& $1$& $0$& $0$ & $1$\\
    \end{tabular}
 % \end{center}
\end{table}

Since there are $ 2^3= 8 $ different neighbourhood states, there are $2^8=256$ different ECA rules. 
It is usual to associate to each ECA $f $ its decimal code $ W $, defined by $ W(f)= f(0,0,0) 2^0 + \dots + f(1,1,1) 2^7$ ~\cite{wolfram1983statistical,wolfram2002}.
This amounts to writing the digits of the transition table $f(0,0,0)$ \dots $ f(1,1,1)$ and converting the binary number to a decimal number.
As an example, Rule 150 corresponds to the array $1,0,0,1,0,1,1,0$, which is equal to 150 in base ten (see Table~\ref{tab:150}).

One can apply two symmetries on the rule, the left-right inversion and the one-zero exchange, which gives 88 classes. The rules with smallest decimal code in a class are called \emph{minimal} CA and it is usual to consider only these rules for studying ECA (see the left column of Table~\ref{tab:fraction}).

%%%%%%%%%%%%%%%%%%%%%%%%%%%%%%%%%%%%%%%%%%%%%%%%%%%%%%%%%%%%%%%%
\subsection{Definition of the regional controllability problem}

\begin{figure}[t]
\centering
\includegraphics[scale=1.15]{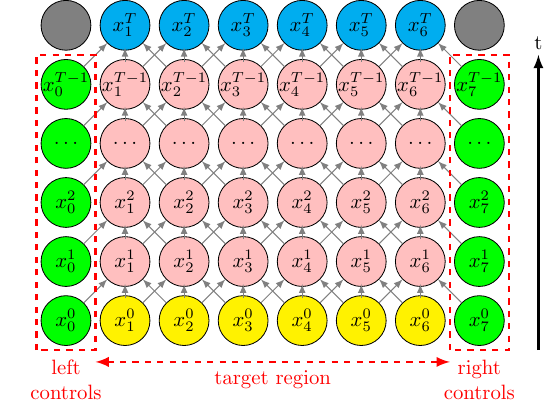}
\caption{General view of the regional controllability problem of one-dimensional CA via boundary actions for $ n= 6 $ controlled cells. The initial and final configurations are respectively in yellow and cyan. Time goes from bottom to top. Colour Online
\label{fig:boundary}}
\end{figure}

In our problem, the definition of a regionally-controllable CA will be set for an arbitrary number of cells, with the goal of determining the asymptotic behaviour. For a given number cells $ n$, we will focus on the region represented by the set of $ n $ cells indexed from 1  to $ n $ and try to control the evolution of this region, without setting in advance the number of steps $ T $. 
We are also interested in looking for the minimal value of $ T(@x,@y) $ for a given pair of configurations $ (@x,@y) $ and the minimal value for a region size $ n $  $$ T(n) = \min \{ T(@x,@y): @x,@y \in \{0,1\}^n \}, $$
where $\{0,1\}^n$ denotes the set of finite words (configurations) of length $n$ over $\{0,1\}$,
  
Given a number of cells $ n  $ and two finite sub-configurations $ @x= (x_1, \dots, x_n) $ and $ @y= (y_1, \dots, y_n)   \in \{0,1\}^n$, our objective is to determine 
%if $ y $ is {\em reachable} from $ @x $, that is, 
if there exists an appropriate control sequence on the boundary cells, namely cell $ 0 $ and cell $ n+1 $, such that the system will evolve from $ @x$ to $ @y $ in  a finite number of time steps (see Fig.~\ref{fig:boundary} for an illustration).

Mathematically, this amounts to saying that:\\

%\begin{definition}
\textit{ A CA  is regionally controllable, if there exists $ N \in \N^+ $ such that for every $ n \geq N$ and for every pair of configurations $ @x=@x^0, @y \in \{0,1\}^n$, there exists $ T > 0 $ and a control vector $ @u=(u^{0},\dots,u^{T-1}) $ where  $u^t=(x_0^{t},x_{n+1}^{t})$ such that 
\[
    @x^T = @y, 
\]
 that is, $ @y $ is reachable from $ @x $ in $ T $ steps, with
  \begin{equation*}
	\forall t \in \{1, \dots, T\}, \forall i \in \{1,\dots, n\},\,\, x^{t+1}_i=f(x^t_{i-1}, x^t_i, x^t_{i+1} ).
  \end{equation*}
}\\

In the case where a rule is not regionally controllable, we are interested in determining the {\em reachability ratio} $ \rr(n) $, that is, the ratio of controllable pairs of initial and final configurations of size $ n $:
\[
    \rr(n) =  \frac{1}{2^{2n}}
        \operatorname{card} \{ (@x,@y) \in \{0,1\}^n \times \{0,1\}^n  : @y \text{ is reachable from }@x \}. 
\]

%%%%%%%%%%%%%%%%%%%%%%%%%%%
\subsection{Peripheral linearity: a sufficient condition to be controllable}

Let us now examine under which conditions elementary cellular automata can be controlled.

To this end, we introduce the notion of a {\em derivative} of a Boolean function $f(x,y)$~\cite{Vichniac1990} as:~
\[
    \frac{\partial f}{\partial x} = f(x\oplus 1, y)\oplus f(x,y),
\]
where $\oplus$ stands for the sum modulo two (equivalent to an exclusive OR).
This definition obeys many standard properties of derivatives, like the chain rule~\cite{bagnoli1992boolean}. It is possible also to define higher-order derivatives. 

By means of Boolean derivatives in zero one can obtain the Ring Sum Expansion of a function~\cite{Wegener1987-kn}:
\begin{equation}\label{eq:rse} 
  \begin{split}
     f(x_0,x_1,x_2) = &D_0 \oplus D_1 x_0 \oplus D_2 x_1 \oplus D_4 x_2 \oplus D_3 x_0 x_1 \oplus \\
         &  D_5 x_0 x_2 \oplus D_6 x_1 x_2 \oplus D_7 x_0 x_1 x_2,
  \end{split}
\end{equation}
where $D_i$ is the derivative of $f$ with respect to the bits that have value 1 in the binary representation of $i$.

For instance 
\[
    D_5 = D_{1,0,1} = \frac{\partial^2 f}{\partial x_0 \partial x_2} (0) = f(1,0,1)\oplus f(1,0,0)\oplus f(0,0,1)\oplus f(0,0,0).
\]

Rules that have all derivatives of order greater than one equal to zero (i.e., $D_3=D_5=D_6=D_7=0$) are  called  {\em affine}, and linear if $f(0,0,0)=0$. 
For instance, Rule 150 is a linear rule since it can be written as
\[
   f^{(150)}(x_0, x_1 , x_2 ) =x_0 \oplus x_1  \oplus x_2,
\] 
while Rule 22, although quite similar to Rule 150 (see Table.~\ref{tab:150}), is not affine:
\[
    f^{(22)}(x_0, x_1, x_2 ) =x_0 \oplus x_1  \oplus x_2 \oplus x_0 x_1 x_2.
\]
The Boolean derivatives in zero and the affinity of all minimal ECA are reported in Table~2 of Ref.~\cite{Dridi2024regional}. 

We can now define the property of {\em linearity with respect to the periphery of the neighbourhood}, which, for ECA, amounts to saying that there exists a function $ g$ such that:
\[
f(x_0,x_1,x_2) = x_0 \oplus g(x_{1}, x_2) \qquad \mathrm{or}\qquad
f(x_0,x_1,x_2) = g(x_0, x_1) \oplus x_2.
\]

In term of Boolean derivatives, a rule is peripherally-linear if $D_1=1$ and $D_3=D_5=D_7=0$ or $D_4=1$ and $D_5=D_6=D_7=0$.  
In the ECA space, the peripherally-linear rules are Rules 15, 30, 45, 60, 90, 105, 106, 150, 154, and 170. 

As shown in Refs.~\cite{bagnoli2016toward,dridi2022kalman}, peripherally-linear rules are fully controllable.

\subsection{Chain transitivity, chain mixing and regional controllability}
Let us recall the definitions of $\varepsilon$-chains, chain transitivity and chain mixing. 

Let $\varepsilon > 0$ and let $@x,@y\in  A^\Z$ where $A$ is the alphabet. An $\varepsilon$-chain ($\varepsilon$-pseudo-orbit) from $@x$ to $@y$ is a finite sequence of configurations  $\{@x^{0},@x^{1},\dots,@x^{t}\}$ with $@x^0=@x$ and $@x^T=@y$, such that for $T> 0$ and $d(F(@x^{t}),@x^{t+1})<\varepsilon$, for $t \in \{0,\dots, T-1\}$,  where the distance $d$ is defined as: \\  

\hspace{.25\linewidth}$
  d(@x,@y)=2^{-\mathrm{min}\left\{|k|:x_{k}\ne y_{k}\right\}} 
$\\
\newline
and  $F:A^\Z \rightarrow A^\Z$ is the global transition function. 

A cellular automaton is chain-transitive if for all $@x,@y \in A^\Z$ and $\varepsilon > 0$ there exists an $\varepsilon$-chain  from $@x$ to $@y$~\cite{kurka2003topological}.  Similarly, a cellular automaton is considered to be chain-mixing if, for any two configurations $@x,@y\in  A^\Z$ and $\varepsilon > 0$ there exists $T>0$  such that for all $t \ge T$, there exists an $\varepsilon$-chain of length $ t $ from $@x$ to $@y$ \cite{kurka2003topological}. 

Recall that regional controllability refers to the ability to steer a dynamical system in a  specific region  from any initial configuration to any desired configuration in that region within a finite time $T$ using appropriate inputs (control).  Meanwhile, chain transitivity means that for any two points in the state space, one can find a sequence of pseudo-orbits (arbitrary small jumps) connecting them. 
The connection between chain transitivity, chain mixing and regional controllability of cellular automata was investigated and a proof has been recently proposed to establish an equivalence between regional controllability for every $n$ ( where $n$ is the size of the controlled region), the chain-transitivity and chain-mixing properties of one-dimensional cellular automata~\cite{dridi2025new}. It has been shown that a CA is regionally controllable for every $n$ if and only if it is chain-transitive and chain-mixing. 

The link between these concepts can be understood through graph theory, by modelling controlled cellular automata as directed graph, where nodes denotes the finite configurations in $A^{n}$ and the arcs represent transitions between them, governed by system dynamics or external inputs. 
In graph terms, the regional controllability concept corresponds to strong connectivity of the state transition graph meaning that any state can be reached from any other state via a directed path between nodes~\cite{dridi2019graph,dridi2019recent}. 
It has been shown that when this connectivity property is verified for every $n$, it is equivalent to chain transitivity and chain mixing~\cite{dridi2025new}. For a finite cellular automaton the relationship also remains valid. 

This means that the value of $ \epsilon $ controls the number of cells $ n $ that we have in our system and that applying the control sufficiently far from the centre cell (say at distance $ n/2$) guarantees that we obtain a valid $\varepsilon$-chain. 
Mathematically, let us consider a configuration $ @x^t $ and its image $ @x^{t+1} = F(@x^t) $. Applying a control on $ @x^{t+1} $ on cells $ -n/2$ and $ n/2 $ transforms this configuration into a configuration $ @y $. It is easy to see that we have $ d(F(@x^t),@y) \leq \frac{1}{2^{n/2}} $, which exactly corresponds to the notion of $\epsilon$-chain for $ \epsilon=\frac{1}{2^{n/2}} $. In other words, we need to apply the control at a larger distance as the value of  $\epsilon$ gets smaller in the $\epsilon$-chain.

%%%%%%%%%%%%%%%%
According to the equivalence presented in  Ref.~\cite{dridi2025new}, 
the methods used to investigate the regional controllability problem can thus also be used to verify the chain transitivity and chain mixing properties.  

\begin{figure}[t]
	\centering
	\includegraphics[width=0.8\linewidth]{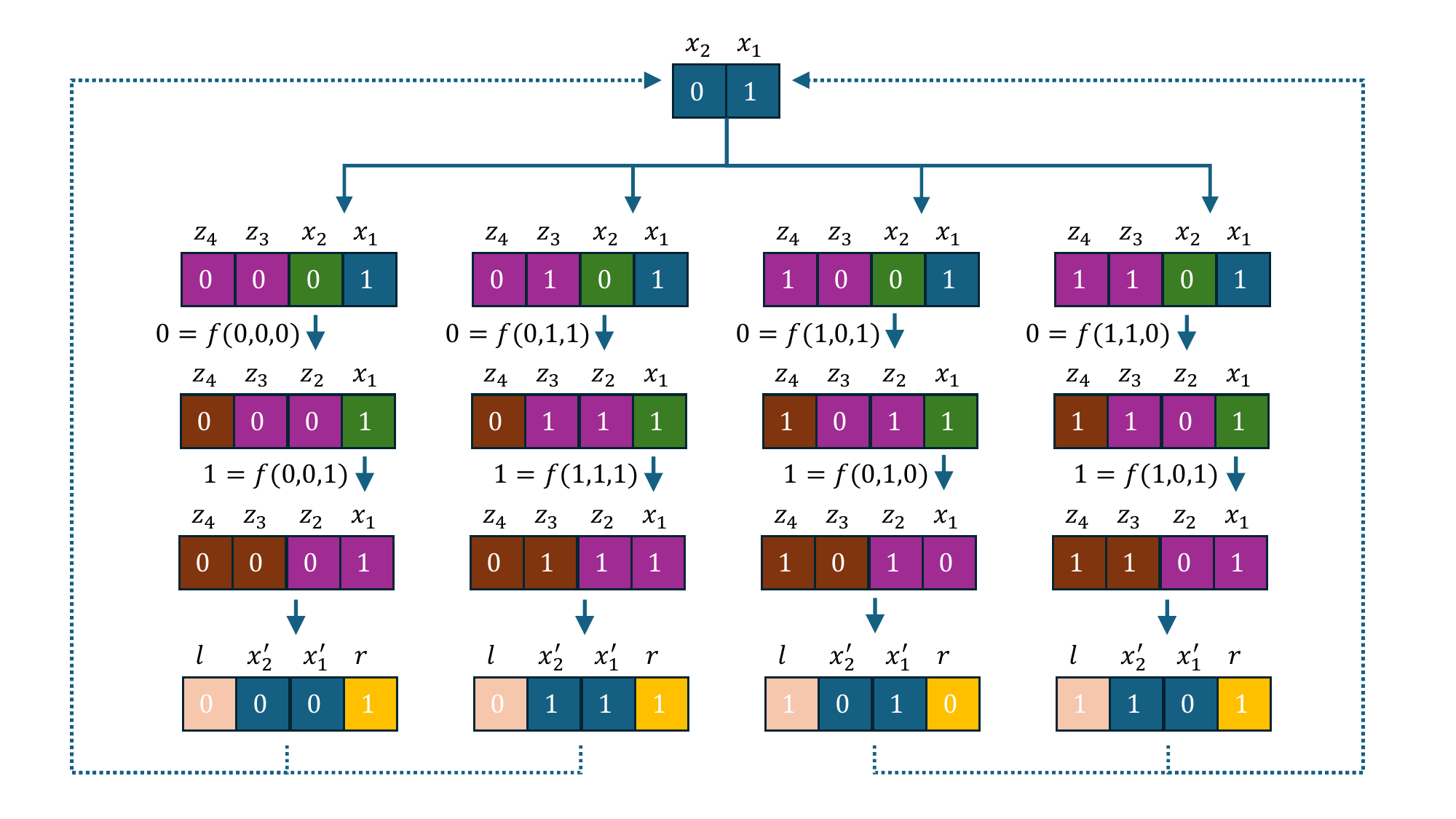}
	\caption{One step of the generation of the preimages (control tree) of configuration $01$ for Rule 150. Since Rule 150 is (doubly) peripherally linear, there is no failures nor additional forking.}
	\label{fig:preimages150}
\end{figure}

%%%%%%%%%%%%%%%%%%%%%%%%%%%%%%%%%%%%%%%%%%%%%%%%%%%%%%%%%%%%%%%%%%%%%%
\section{Modelling the control problem as a SAT problem}\label{sec:3}

\begin{figure}[t]
	\centering
	\includegraphics[scale=0.95]{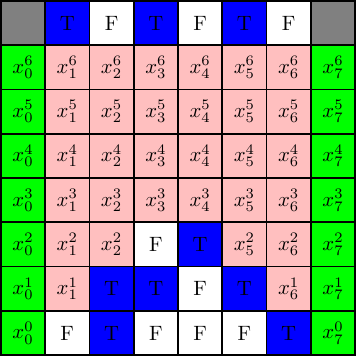}
	\hspace{.05\linewidth}
	\includegraphics[scale=0.95]{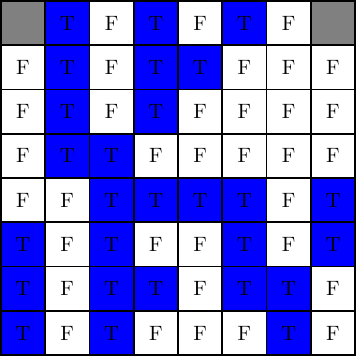}
	\caption{(left) Modelling of the controllability problem with Rule 30, initial condition \texttt{010001}, final condition \texttt{101010} and $ T = 7 $ time steps. Time goes from bottom to top. The control variables are shown in green and the controlled variables in pink. The values of two control variables $ x^7_0 $ and $ x^7_7 $ are irrelevant and are left in grey.
		(right) A solution that was found by the SAT solver. White and blue respectively correspond to \texttt{0} and \texttt{1}, which are respectively associated to \texttt{False} (F) and \texttt{True} (T). Color online}
	\label{fig:satcontrol}
\end{figure}

We can now investigate whether the control problem is solvable for non-peripherally linear rules.
In this section, we reformulate the regional controllability of cellular automata problem as a SAT problem.

%%%%%%%%%%%%%%%%%%%%%%%%%%%%%%%%%%
\subsection{Modelling the problem}

Recall that we have $ n $ cells in the target region with indices ranging from 1 to $ n $, and two external cells with index $ 0 $ and $ n+1$ as the border controls. In the CA world, $ x_i^t \in \{0,1\} $ describe the state of cell $ i \in \{1, n\} $ at time $ t $ and $ x_0^t$ and $ x^t_{n+1}$ respectively describe the state of left and right controls at time $ t $.
To go to the SAT universe, we map each binary value $ x_i^t $ to a Boolean value $ b_i^t $, with $ t \in \N $ and $ i \in \{0, \dots, n\}$ such that $ b_i^t =\BB(x_i^t) $, where $\BB$ is the function from $ \{0,1\} $ to $ \{\texttt{False},\texttt{True}\} $ such that $ \BB(0)= \texttt{False} $ and $ \BB(1)= \texttt{True}$.

Recall that we fix the values $ @x=(x_1,\dots,x_n) $ and that we want to reach $ @y=(y_1,\dots,y_n) $ in $ T $ time steps.
This is equivalent to saying that there exists two sequences $ (@x_0^t)_{t\in \{0,\dots,T-1\}} $ and $ (@x_{n+1}^t)_{t\in\{0,\dots,T-1\}} $  such that:

\begin{equation}
  \forall t \in \{0, \dots, T-1\}, \forall i  \in \{1,\dots,n\},\,\,
    x_i^{t+1} = f(x_{i-1}^t, x_{i}^t, x_{i+1}^t) \label{Eq:evol}.
\end{equation}
and $ x_i^T = y_i $ for all $ i \in \{1,\dots, n\}$, that is, $ @y $ is reachable from $ @x $ in $ T $ steps.

\smallskip
We now need to transform this mathematical relationship into a Boolean formula. Such a formula is usually expressed as a combination of variables, or atoms, and operators. A {\em clause} is a special case of formula where atoms and their negations are combined using only \texttt{OR} operators, for instance $ f(a,b,c)= a \vee \neg b \vee c$.
A formula is in {\em conjunctive normal form} (CNF) if it is expressed as a conjunction of clauses, for instance: 
$ f(a,b,c,d)= (a \vee \neg b ) \wedge (b \vee \neg c \vee \neg d)$.
SAT solvers take a CNF as an input and try to compute an assignment of the variables that makes the formula true, or, when this is not possible, try to find a proof that there this no such assignment.

To encode the CNF that verifies the regional controllability  problem, we proceed in two steps (see Fig.~\ref{fig:satcontrol}).

a) For each $ i $ and $ t $, we encode the condition $ x_i^{t+1} = f(x_{i-1}^t, x_{i}^t, x_{i+1}^t)$ as a CNF composed of eight clauses which
use the four variables $ b_i^{t+1}$, $ b_{i-1}^t$, $b_{i}^t$, and $ b_{i+1}^t $.

Indeed, each neighbourhood state will generate a clause:
\[ 
 \begin{split}
    (x_{i-1}^t, x_{i}^t, x_{i+1}^t) &= (0,0,0) \implies x_i^{t+1} = f(0,0,0) ,  \\
    (x_{i-1}^t, x_{i}^t, x_{i+1}^t) &= (0,0,1) \implies x_i^{t+1} = f(0,0,1) , \dots,\\
    (x_{i-1}^t, x_{i}^t, x_{i+1}^t) &= (1,1,1) \implies x_i^{t+1} = f(1,1,1) .
 \end{split}
\]
As $ A \implies B $ is equivalent to $ \neg A \vee B $, the conditions becomes
\[
    x_{i-1}^t \neq 0 \vee  x_{i}^t \neq 0 \vee x_{i+1}^t \neq 0 \vee x_i^{t+1} = f(0,0,0), \dots, 
\]
that is,
\[ 
    x_{i-1}^t =1 \vee x_{i}^t =1 \vee x_{i+1}^t = 1 \vee x_i^{t+1} = f(0,0,0), \dots 
\]
which translates into
\[
    b_{i-1}^t \vee b_{i}^t  \vee b_{i+1}^t \vee C[b_i^{t+1}, f(0,0,0)] \dots, 
\]
where $ C [b,1] = b $ and $ C[b,0]= \neg b $.

We thus translate the $ n \cdot T $ transitions described in Eq.~\eqref{Eq:evol} into $ 8 \cdot n \cdot T $ clauses.

\smallskip
b) In a second step, we encode the initial and final conditions $ (b_i^0) $ and $ (b_i^T) $ for  $ i \in \{1, \dots, n\} $ and let the control variables free of any constraint.

By combining the conditions obtained in step a) and step b), we obtain a CNF formula $ \Phi $ with $ 8 n T + 2n $ clauses and such that $ \Phi $ is satisfiable if and only if  $ @y $ is reachable from $ @x $ in $ T $ time steps.

%%%%%%%%%%%%%%%%%%%%%%%%%%%%%%%%
\subsection{Experimental results}
\label{sec:expResults}

\newcommand{\lw}{\linewidth}
\newcommand{\xcmd}[7]{ #1 & #2 & #3 & #4 & #5 & #6}
\begin{table}[t]
\caption{Table showing the controllability statistics for the 88 ECA for $ n =40 $, $ T= 100 $ and 100 random pairs of initial and final configurations. columns : $\tilde\rr$ is the average reachability ; time is the cumulative CPU time ; R, C, D respectively represent the cumulative number of restarts, conflicts and decisions of the SAT solver.}
%\tiny
\begin{tabular}{ | p{.05\lw} p{.03\lw} p{.08\lw} | p{.05\lw} p{.07\lw} p{.12\lw} | | p{.05\lw} p{.03\lw} p{.08\lw} | p{.05\lw} p{.07\lw} p{.12\lw} |}
    \hline
    ECA & $\tilde\rr$ & time &  R  &   C &   D  & ECA & $\tilde\rr$ & time & R  &   C &   D \\
    \hline
    \xcmd{0}{0.00}{1.874}{0}{0}{0}{377623} & \xcmd{1}{0.00}{1.870}{0}{0}{0}{74950} \\
    \xcmd{2}{0.00}{1.866}{0}{0}{0}{53801} & \xcmd{3}{0.01}{1.838}{1}{0}{1}{62483} \\
    \xcmd{4}{0.00}{1.834}{0}{0}{0}{70341} & \xcmd{5}{0.00}{1.864}{0}{0}{0}{80056} \\
    \xcmd{6}{0.00}{1.922}{0}{0}{0}{50633} & \xcmd{7}{0.00}{1.765}{3}{39}{143}{60936} \\
    \xcmd{8}{0.00}{1.808}{0}{0}{0}{52392} & \xcmd{9}{0.00}{1.942}{0}{0}{0}{52498} \\
    \xcmd{10}{0.00}{1.985}{0}{0}{0}{58789} & \xcmd{11}{0.00}{1.870}{0}{0}{0}{52177} \\
    \xcmd{12}{0.00}{1.757}{0}{0}{0}{68419} & \xcmd{13}{0.00}{2.073}{0}{0}{0}{75919} \\
    \xcmd{14}{0.01}{1.825}{1}{14}{451}{57757} & \xcmd{15}{1.00}{2.190}{100}{0}{100}{168000} \\
    \xcmd{18}{0.00}{1.949}{0}{0}{0}{55047} & \xcmd{19}{0.00}{1.867}{0}{0}{0}{69489} \\
    \xcmd{22}{0.07}{6.727}{929}{281123}{780831}{13879974} & \xcmd{23}{0.00}{1.852}{0}{0}{0}{51415} \\
    \xcmd{24}{0.00}{2.018}{0}{0}{0}{49467} & \xcmd{25}{0.00}{2.182}{11}{151}{838}{59552} \\
    \xcmd{26}{0.05}{2.085}{42}{4554}{28485}{454868} & \xcmd{27}{0.00}{1.916}{0}{0}{0}{51904} \\
    \xcmd{28}{0.00}{1.930}{0}{0}{0}{50889} & \xcmd{29}{0.00}{2.140}{0}{0}{0}{54568} \\
    \xcmd{30}{1.00}{3.158}{100}{0}{100}{60826} & \xcmd{32}{0.00}{2.095}{0}{0}{0}{57121} \\
    \xcmd{33}{0.00}{2.202}{0}{0}{0}{52552} & \xcmd{34}{0.00}{2.008}{0}{0}{0}{54310} \\
    \xcmd{35}{0.00}{1.987}{0}{0}{0}{52608} & \xcmd{36}{0.00}{2.161}{0}{0}{0}{50912} \\
    \xcmd{37}{0.00}{2.142}{23}{163}{495}{59991} & \xcmd{38}{0.00}{2.092}{0}{0}{0}{50744} \\
    \xcmd{40}{0.00}{2.072}{0}{0}{0}{60236} & \xcmd{41}{0.01}{3.488}{221}{72085}{177628}{4235321} \\
    \xcmd{42}{0.04}{1.937}{4}{0}{4}{64161} & \xcmd{43}{0.01}{1.849}{1}{41}{2190}{58431} \\
    \xcmd{44}{0.00}{1.795}{0}{0}{0}{51963} & \xcmd{45}{1.00}{2.506}{100}{0}{100}{62163} \\
    \xcmd{46}{0.00}{1.867}{0}{0}{0}{49610} & \xcmd{50}{0.00}{1.923}{0}{0}{0}{58289} \\
    \xcmd{51}{0.00}{1.760}{0}{0}{0}{72265} & \xcmd{54}{0.00}{1.947}{3}{148}{1023}{57370} \\
    \xcmd{56}{0.00}{1.813}{0}{0}{0}{49960} & \xcmd{57}{0.00}{1.870}{0}{0}{0}{52218} \\
    \xcmd{58}{0.00}{1.898}{0}{0}{0}{52029} & \xcmd{60}{1.00}{2.093}{100}{0}{100}{86000} \\
    \xcmd{62}{0.00}{2.061}{10}{555}{1883}{89468} & \xcmd{72}{0.00}{1.722}{0}{0}{0}{64893} \\
    \xcmd{73}{0.00}{1.809}{0}{0}{0}{54992} & \xcmd{74}{0.00}{1.866}{3}{18}{37}{52060} \\
    \xcmd{76}{0.00}{1.763}{0}{0}{0}{71130} & \xcmd{77}{0.00}{1.790}{0}{0}{0}{51260} \\
    \xcmd{78}{0.00}{1.759}{0}{0}{0}{52732} & \xcmd{90}{1.00}{2.117}{100}{0}{100}{46000} \\
    \xcmd{94}{0.00}{1.777}{0}{0}{0}{57453} & \xcmd{104}{0.00}{1.811}{0}{0}{0}{61062} \\
    \xcmd{105}{1.00}{2.485}{100}{0}{100}{46000} & \xcmd{106}{1.00}{2.553}{100}{0}{100}{63060} \\
    \xcmd{108}{0.00}{1.914}{0}{0}{0}{52426} & \xcmd{110}{0.04}{7.009}{667}{222617}{499441}{17298571} \\
    \xcmd{122}{0.00}{1.874}{0}{0}{0}{51215} & \xcmd{126}{0.03}{3.445}{249}{62888}{183969}{4687251} \\
    \xcmd{128}{0.00}{1.795}{0}{0}{0}{193238} & \xcmd{130}{0.00}{1.864}{0}{0}{0}{125813} \\
    \xcmd{132}{0.00}{1.800}{0}{0}{0}{69357} & \xcmd{134}{0.00}{1.804}{5}{93}{394}{55391} \\
    \xcmd{136}{0.00}{1.852}{0}{0}{0}{55748} & \xcmd{138}{0.00}{1.783}{0}{0}{0}{59102} \\
    \xcmd{140}{0.00}{1.729}{0}{0}{0}{71952} & \xcmd{142}{0.00}{1.776}{0}{0}{0}{50822} \\
    \xcmd{146}{0.01}{2.093}{39}{5636}{25835}{405800} & \xcmd{150}{1.00}{2.498}{100}{0}{100}{46000} \\
    \xcmd{152}{0.00}{1.842}{0}{0}{0}{53363} & \xcmd{154}{1.00}{2.608}{100}{0}{100}{72859} \\
    \xcmd{156}{0.00}{1.899}{0}{0}{0}{55236} & \xcmd{160}{0.00}{1.805}{0}{0}{0}{178336} \\
    \xcmd{162}{0.00}{1.907}{0}{0}{0}{60126} & \xcmd{164}{0.00}{1.810}{0}{0}{0}{57425} \\
    \xcmd{168}{0.00}{1.864}{0}{0}{0}{60767} & \xcmd{170}{1.00}{2.095}{100}{0}{100}{168000} \\
    \xcmd{172}{0.00}{1.787}{0}{0}{0}{53371} & \xcmd{178}{0.00}{1.773}{0}{0}{0}{50993} \\
    \xcmd{184}{0.00}{1.765}{0}{0}{0}{50385} & \xcmd{200}{0.00}{1.780}{0}{0}{0}{68599} \\
    \xcmd{204}{0.00}{1.814}{0}{0}{0}{71348} & \xcmd{232}{0.00}{1.832}{0}{0}{0}{52025} \\
    \hline
\end{tabular}
\label{tab:expSATsolve}
\end{table}

We tested the solving abilities of the SAT solvers with the \texttt{minisat} solver~\cite{minisat}.
We used a simple script to generate the formula seen above in the form of a CNF expressed in the popular DIMACS format.
For a fixed value of $ n $ and $ T $, we randomly generated a pair of initial and final configurations $ @x, @y \in \{0,1\}^n $ and asked the solver whether $ @y $ was reachable from $ @x $ with a control sequence of length $ T $.

Table~\ref{tab:expSATsolve} presents the results for the 88 minimal ECA for the setting $ n = 40 $, $ T= 100 $ and a sample of 100 random pairs of initial and final configurations. The second column ($\tilde\rr$) presents the ratio of cases where a solution was found, that is, a rough estimate of the reachability ratio $ \rr(n)$. We observe that $\tilde\rr$ is either equal to 1 or is rather small (less than 10\%). The case where  $\tilde\rr$ equals one corresponds exactly to the ten peripherally-linear rules 15, 30, 45, 60, 90, 105, 106, 150, 154, and 170. 
For the other rules, we issue the following conjecture :\\

\textit{The reachability ratio $ \rr(n) $ tends to zero when the number of cells $ n $ tends to infinity for all the ECA rules which are not peripherally-linear.}\\

This conjecture is supported by the data shown on Table~\ref{tab:scaling}, where one can clearly see that the estimated reachability ratio $ \tilde\rr(n) $ rapidly decreases with $ n $, and becomes of the order of a percent when $ n $ is equal to 50. It is an open question to obtain more precise scaling laws with $ n $, and a good estimate of the time needed to reach a configuration from another.

Table~\ref{tab:expSATsolve} also presents the CPU time that is used to compute the presence or the absence of a solution, and the number of restarts, conflicts and  decisions that were taken by the SAT solver. These figures are only presented to furnish a rough estimate of the difficulty of analysis that was met for each rule. It should be noted that we did not encounter any case where the solver was unable to provide an answer, either positive or negative and that the answer was given quite rapidly ({$\approx$} 20\ ms per solution on average). 

It is also remarkable that the among the rules for which the number of restarts is important, one finds the famous Rule 110, which is known to be Turing-universal. Of course, there is no direct correlation between the computational complexity of a rule and the difficulty to decide its controllability, but we can safely state that the rules which exhibit the richest panel of behaviours (gliders, collisions generating other gliders, etc.) are among the most difficult ones to analyse by the SAT solvers.

In order to gain insights on the relationships between regional controllability and the difficulty to find a solution or to prove that there is none, we now explore a technique of tree-building in order to find the shortest paths between two configurations.

%%%%%%%%%%%%%%%%%%%%%%%%%%%%%%%%%%%%%%%%%%%%%%%%%%%%%

%%%%%%%%%%%%% SCALING with n TAB %%%%%%%%%%%%%%%%
\newcommand{\ycmd}[7]{#1 & #2 & #3 & #4 & #5 & #6 & #7}
\begin{table}[t]
\caption{Estimation of the reachability ratio $ \rr(n) $ for values of $ n $ ranging from 10 to 50, for Rule 22 and Rule 110 and different values of $ T $. Statistics on 100 random pairs of  initial and final random configurations.}
\label{tab:scaling}
\begin{tabular}{|p{0.03\lw} c | c c c c c|}
\multicolumn{2}{c}{} &  \multicolumn{5}{c}{$n$}\\
\hline
\ycmd{ ECA }{ T }{ 10 }{ 20 }{ 30 }{ 40 }{ 50 }\\
\hline
\ycmd{ 22 }{ 100 }{ 0.83 }{ 0.38 }{ 0.15 }{ 0.03 }{ 0.01 }\\
\ycmd{ 22 }{ 200 }{ 0.84 }{ 0.46 }{ 0.11 }{ 0.04 }{ 0.01 }\\
\ycmd{ 22 }{ 400 }{ 0.89 }{ 0.43 }{ 0.17 }{ 0.03 }{ 0.01 }\\
\hline
\ycmd{ 110 }{ 100 }{ 0.72 }{ 0.38 }{ 0.17 }{ 0.03 }{ 0.01 }\\
\ycmd{ 110 }{ 200 }{ 0.75 }{ 0.36 }{ 0.15 }{ 0.07 }{ 0.00 }\\
\ycmd{ 110 }{ 400 }{ 0.75 }{ 0.31 }{ 0.13 }{ 0.06 }{ 0.01 }\\
\hline
\end{tabular}
\end{table}

\begin{figure}[t]
	\centering
	\includegraphics[width=0.8\linewidth]{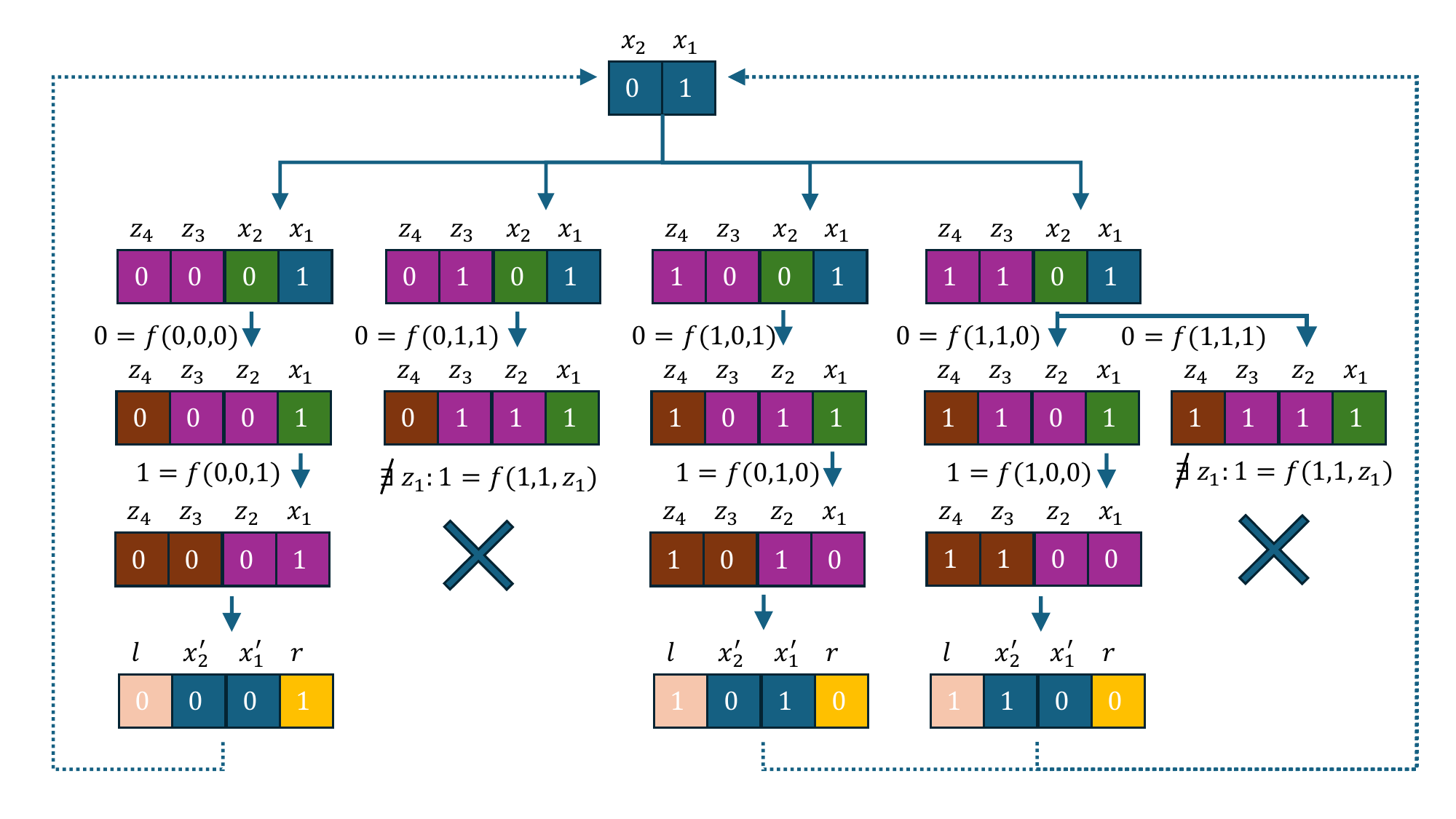}
	\caption{One step of the generation of the preimages (control tree) of configuration $01$ for Rule 22. Since Rule 22 is not peripherally-linear, some configurations do not have preimages and others have more than one preimage.}
	\label{fig:preimages22}
\end{figure}

%%%%%%%%%%%%%%%%%%%%%%%%%%%%%%%%%%%%%%%%%%%%%%%%%
\section{Generating preimages }\label{sec:4}

Let us sketch the idea of generating the preimage   $(z_0, z_1, z_2, \dots, z_n, z_{n+1})$   of a given configuration $(x_1, x_2, \dots x_n)$, for all possible controls $z_0$ and $z_{n+1}$. This method will be exploited in the next section to examine the control tree.

The construction of the preimage is performed one site after the other. We start by inserting the two sites $z_0$ ad $z_1$. 
Since there are $n+2$ items in the preimage   $(z_0, z_1, z_2, \dots, z_n, z_{n+1})$, and only $n$ in $(x_1, x_2, \dots, x_n)$, 
we have to  iterate over all the four values of $z_0$ and $z_1$.

We can then select the value(s) of $z_2$ such that
\[
x_1 = f(z_0,z_1, z_2).
\]
It may happen that there are no values allowed, just one or two. 

In the first case, we abort the reconstruction and pass to the next pair of $z_0$ and $z_1$ values, if available, or declare that there is no preimage. In the second case, we can proceed with the neighbouring site ($z_3$). In the third case, we have to fork  the procedure for the two values of $z_2$ and continue to the next site ($z_3$). 
This procedure continues until we generate $z_{n+1}$. 

Instead of scanning the whole look-up table (which can be costly for large neighbourhoods), one can take profit of the ring sum expansion, Eq.~\eqref{eq:rse} 
\[
    \begin{split}  
x_i =  &D_0 \oplus D_1 z_{i-1} \oplus D_2 z_{i} \oplus D_4 z_{i+1} \oplus D_3 z_{i-1} z_{i} \oplus \\
 & D_5 z_{i-1} z_{i+1} \oplus D_6 z_{i} z_{i+1} \oplus D_7 z_{i-1} z_{i} z_{i+1},
 \end{split}
\]
to calculate directly $ z_{i+1}$ knowing $ z_{i-1}$ and $z_i$, using the formula:
\[
    \begin{split}  
    z_{i+1}(D_4 & \oplus D_5 z_{i-1} \oplus D_6 z_{i} \oplus D_7  z_{i-1} z_{i}) = \\
     & x_i \oplus D_0 \oplus D_1 z_{i-1} \oplus D_2 z_{i} \oplus D_3 z_{i-1} z_{i},
     \end{split}
\]
as illustrated in Fig.~\ref{fig:preimages22} for Rule 22. 

If $D_4 \oplus D_5 z_{i-1} \oplus D_6 z_{i} \oplus D_7  z_{i-1} z_{i}=1$ then $z_{i+1} = x_i \oplus D_0 \oplus D_1 z_{i-1} \oplus D_2 z_{i} \oplus D_3 z_{i-1} z_{i}$, otherwise, if $x_i = D_0 \oplus D_1 z_{i-1} \oplus D_2 z_{i} \oplus D_3 z_{i-1} z_{i}$,  one has to fork for $z_{i-1}=0,1$, else the recursion fails.

An example of this procedure is reported in Figure~\ref{fig:preimages150} for a peripherally-linear rule (Rule 150) and in Figure~\ref{fig:preimages22} for a nonlinear rule (Rule 22).

Let us illustrate in detail the operation: the evolution of a rule which is right-peripheral, i.e., 
\[
x' =  g(x_{-}, x) \oplus x_{+} 
\]
 can be inverted, giving
\[%begin{equation}\label{eq:pl}
 x_{+} = x' \oplus g(x_{-}, x),
\]%end{equation}
and in this case it is evident that the preimage is always unique. 

For instance, for Rule 150
\[
 x' = x_{-}\oplus x\oplus x_+ \qquad \Rightarrow \qquad x_+ = x'\oplus  x\oplus x_-.
\]

On the other side, for nonlinear rules like Rule 22
\[
 x' = x_{-}\oplus x\oplus x_+ \oplus x_-xx_+ \qquad \Rightarrow \qquad x_+ (1\oplus x_{-}x) = x'\oplus x_{-}\oplus x_+,
\]
we have three possibilities: 
if $1\oplus x_-x=1$; one can obtain $x_+$ uniquely, otherwise, either we have zero or two possibilities. In the example of Rule 22, $1\oplus x_-x=0$ implies $x_-=x=1$ and looking in the Look-up table (Table~\ref{tab:150}) we can only have $x'=0$, so, if this is the case, we have either both values $x_+=0$ and $x_+=1$ (if $x'=0$), or none. 

%%%%%%%%%%%%%%%%%%%%%%%

\begin{figure}[t]
     \centering
     \includegraphics[width=0.8\linewidth]{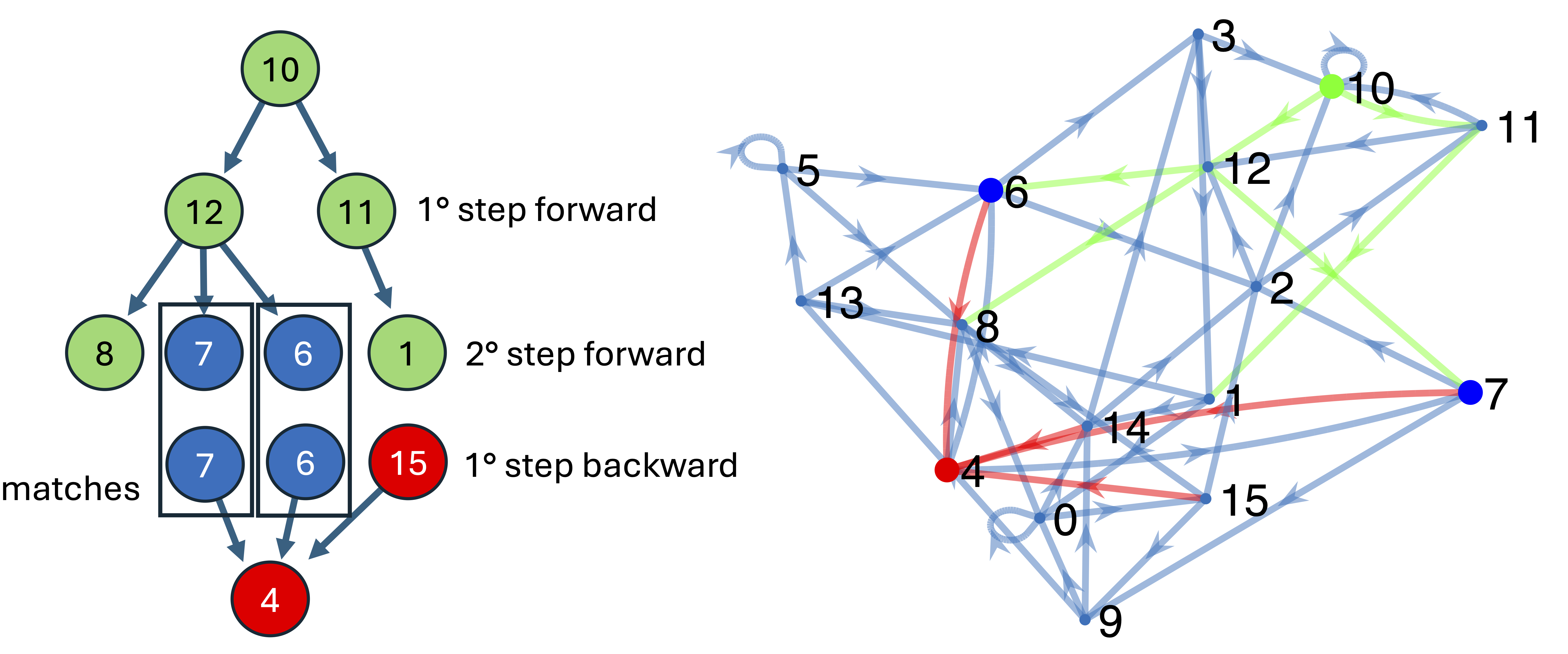}
     \caption{The two minimal control paths from configuration 10 to configuration 4 for Rule 30 and $n=3$. The forward  edges are in green and the backward ones in red. The connecting configurations  6 and 7 are marked in blue. To the left the full control graph (controls values are not shown, for each link there could be more than one control possible).}
     \label{fig:tree}
 \end{figure}

%%%%%%%%%%%%%%%%%%%%%%%%%%%%%%%%%%%%%%%%%%%%%%
\section{An alternative approach:Finding the shortest path in the control tree}\label{sec:tree}

We present here an alternative method to test the reachability of $ @y $ from $ @x $: we simultaneously explore both the tree of images starting from $@x$, with all the possible controls, and the  tree of preimages starting from $@y$, until these two trees meet in at least one configuration, thus indicating the shortest control that transforms $ @x $ into $ @y $.

Finding a control $@l=(x_0^0, x_0^1,\dots x_0^{T-1})$ and $@r=(x_{n+1}^0, x_{n+1}^1, \dots, x_{n+1}^{T-1})$ driving the system from configuration $@x=@x^0=(x^0_1,x^0_2\dots,x^0_n)$ at time $t=0$ to configuration $@y=@x^T=(x^T_1,x^T_2\dots,x^T_n)$ at time $T$ can be done by looking for the shortest path in the tree generated by all possible pairs of values for the control at each time step, forward in time starting from $@x$, or in the tree of all possible preimages backward in time starting from configuration $@y$. 

Going forward in time, each node can in principle generate four branches, for the four possible values of the left-right pair, but it may happen that more than one control pair gives the same configuration. Going backward in time, it may also happen that a given control pair gives no configuration, or, as we have seen, more branches have to be followed while generating preimages. 

\begin{figure}[t]
    \centering
    \includegraphics[width=0.8\linewidth]{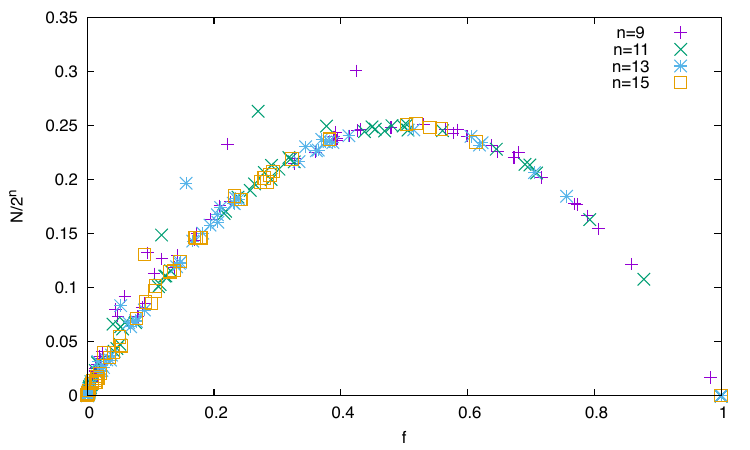}
    \caption{Average number of nodes $N$ visited to determine the absence of a control for the 88 minimal rules, scaled by the the total number of configurations ($2^n$) vs. the reachability ratio $\rr$. Peripherally-linear rules have $\rr=1$ and $N=0$. Data from Table~\ref{tab:fraction}.
        \label{fig:fraction}}
\end{figure}

Depending on the branching of the trees, it may be more efficient to generate more forward or backward levels.  However, an acceptable strategy consists in generating one level at a time for the forward and the backward tree, until the same configuration appears at the end of both, marking a shortest path from $@x$ to $@y$.  This procedure is indicated schematically in Figure~\ref{fig:tree}. 

An estimation of the reachability ratio $\rr$ for different lengths and all minimal elementary cellular automaton is reported in Table~\ref{tab:fraction}. One can see that  peripherally-linear ECA are always fully controllable, while the fraction of controllable pairs for the other ECA diminishes with the length of the configuration, therefore indicating that the boundary control is almost impossible for large lengths for  non-peripherally-linear rules. This confirms the  observations made in Section~\ref{sec:expResults} that for the ECA that are not peripherally-linear, the controllability of randomly chosen pairs tends to zero as the size of the system grows.

This technique can help us compute an estimation of the reachability ratio $\rr(n)$. We can also estimate the average complexity of finding all controls by counting the average number of nodes explored in the control tree before declaring that there is no possible control. This complexity is zero for peripherally-linear rules (since they are always controllable), but also for CA like Rule 0, which have only one possible configuration in their image (all-zero). This also holds for all the rules that have a reachability ratio very near to zero, see Table~\ref{tab:fraction}.

In Figure~\ref{fig:fraction} we show that there is a nice relationship between the reachability ratio~$\rr$ and the average complexity, scaled by the total number of configurations ($2^n$). This behaviour is reminiscent of the complexity phase transitions of SAT problems~\cite{Hayes,WhereReallyHardProblems,Kirkpatrick1994,PhaseTransitions}.

%%%%%%%%%%%%%%%%%%%%%
\section{Conclusions}\label{sec:5}

In control theory, systems are classically defined by a set of differential equations and the control is applied with a feedback system. In this work, we extended this framework by tackling the question of how to control a {\em spatially-extended discrete dynamical system}, namely an elementary cellular automaton, by changing the state of only two boundary cells at each time step  to influence a region of the entire domain.  

We showed that the problem could adequately by translated into a Boolean CNF formula and fed to a SAT solver to effectively obtain a solution when the control is indeed possible, or a proof that it is impossible when this the case. From a concrete point of view, the control sequence is obtained rapidly, even for a system of the order of a hundred cells and a few hundred time steps. Interestingly enough, we observed that the difficulty to find a solution was somehow related to the computational complexity of the rules. As an alternative, we showed that one could also search for the shortest path between an initial and a final configuration by progressively extending trees in the directed graph of the transitions between configurations.

From a theoretical point view, we showed the equivalence of regional controllability for every $n$   with being chain-transitive and chain-mixing  as presented in Ref.~\cite{dridi2025new}. 

This means that our techniques may also apply to verify such properties in various other cases. More generally, a logical next step would be to explore how our work extends to cellular automata with  higher dimensions, a greater number of states, or even to non-uniform rules (applying different local rules in a SAT solver and in the search of the shortest path can be done directly without any additional effort).

We also observed that in the ECA that are not peripherally-linear, the probability to reach a final configuration from an initial configuration vanishes when the system's size grows and when the configurations are chosen randomly. It would be interesting to prove this property formally. In fact, a more precise description would quantify how the size of the communication classes of the transition graph scale as a function of the system's size. The techniques that were applied for fully asynchronous ECA~\cite{RFD24}  may also apply here and the use of SAT solvers may be of great help to derive such formal proofs of reachability.

\newpage

\begin{longtable}{c|c|c||c|c||c|c||c|c}
\caption{\normalsize Numerical estimation $\tilde\rr$ of the reachability ratio and number $N$ of nodes to be visited to check for the absence of control for all minimal ECA $R$ and various configuration sizes $n$. The data are obtained sampling over $10^4$ random pairs of configurations.} \label{tab:fraction} 
\\
{\normalsize $R$} & \multicolumn{2}{c||}{\normalsize $n=9$}&
\multicolumn{2}{c||}{\normalsize $n=11$}&
\multicolumn{2}{c||}{\normalsize $n=13$}&
\multicolumn{2}{c}{\normalsize $n=15$}\\
\hline
 & $\tilde\rr$ & N & $\tilde\rr$ &N& $\tilde\rr$ &N &  $\tilde\rr$ &N\\
 \hline
0&0.002&0.9956&0.0011&0.9985&0&0.9997&0&0.9999\\
1&0.0112&9.8817&0.0021&10.2039&0.0005&10.2994&0.0003&9.1978\\
2&0.0801&37.6393&0.0426&84.2125&0.0208&185.0444&0.0137&400.4257\\
3&0.3936&121.053&0.2907&436.7135&0.2303&1459.9432&0.1783&4800.218\\
4&0.0026&2.5131&0.0007&2.7917&0.0002&2.7316&0.0001&3.3823\\
5&0.0132&12.9694&0.0031&14.1058&0.0006&13.7366&0&14.5212\\
6&0.3918&124.9131&0.2638&401.5478&0.1666&1172.6215&0.107&3171.2119\\
7&0.2093&90.0002&0.1223&226.8278&0.0655&552.3652&0.0417&1302.8289\\
8&0.0035&3.6936&0.0013&4.0129&0.0003&4.1582&0&4.4093\\
9&0.3893&122.727&0.2378&376.5136&0.141&978.621&0.0774&2346.7942\\
10&0.195&83.5645&0.1316&236.9626&0.0902&649.5169&0.0507&1775.1363\\
11&0.327&112.401&0.2183&349.5516&0.1468&1011.7967&0.0917&2837.0574\\
12&0.003&2.6059&0.0007&2.7152&0.0004&2.8586&0&3.0346\\
13&0.0156&14.6131&0.0038&17.273&0.0014&19.732&0.0004&20.4391\\
14&0.4133&123.18&0.2794&423.1061&0.1938&1293.3974&0.1304&3763.2499\\
15&1&0&1&0&1&0&1&0\\
18&0.4262&125.7239&0.3267&443.4179&0.234&1497.6062&0.1713&4780.7041\\
19&0.0083&7.4527&0.0018&7.93&0.0004&7.9429&0.0002&8.3817\\
22&0.8586&62.3896&0.7928&333.8604&0.7057&1692.3004&0.6131&7699.2784\\
23&0.0126&12.1348&0.0031&13.4406&0.0008&14.4739&0.0005&14.9865\\
24&0.0911&43.8604&0.0515&97.4022&0.0257&213.4277&0.0142&461.6439\\
25&0.7168&103.373&0.5606&502.6101&0.4128&1971.0832&0.2882&6698.8188\\
26&0.7686&90.9939&0.6906&438.8354&0.6189&1900.8206&0.5183&8258.4423\\
27&0.5841&126.0192&0.4693&501.9368&0.3645&1862.9912&0.2838&6487.5229\\
28&0.0235&20.7356&0.0058&26.5314&0.0024&29.8961&0.0003&34.9573\\
29&0.0156&13.5261&0.0039&14.127&0.0013&14.6097&0.0002&14.692\\
30&1&0&1&0&1&0&1&0\\
32&0.0062&5.1255&0.0029&5.1161&0.0003&5.5633&0&5.319\\
33&0.0447&40.8937&0.0128&47.7303&0.0028&50.5435&0.001&50.9947\\
34&0.1698&73.7404&0.1125&206.6842&0.0776&562.5955&0.0541&1510.5548\\
35&0.5773&124.4786&0.4492&509.7556&0.3446&1888.0347&0.2731&6487.2539\\
36&0.0064&6.3585&0.0022&7.1627&0.0005&7.0318&0.0003&7.1574\\
37&0.9829&8.5978&0.8781&220.7768&0.7562&1512.4127&0.5589&8097.3136\\
38&0.3943&121.9057&0.2901&410.0903&0.2061&1316.2389&0.1464&4065.0292\\
40&0.0201&17.7929&0.0063&20.9039&0.0011&21.9891&0.0003&28.0374\\
41&0.8068&79.0688&0.6983&437.369&0.6055&1966.9365&0.5038&8234.82\\
42&0.5302&128.4775&0.4543&505.6233&0.3875&1920.5669&0.323&7182.0698\\
43&0.6385&118.688&0.506&503.6215&0.37&1943.5089&0.2789&6608.7907\\
44&0.0145&13.4614&0.0037&15.1588&0.0014&15.5487&0.0002&16.7039\\
45&1&0&1&0&1&0&1&0\\
46&0.1373&60.7865&0.0762&138.4218&0.0383&307.8265&0.0203&671.4806\\
50&0.0126&12.9995&0.0046&14.382&0.0009&15.2791&0.0003&16.3391\\
51&0.004&1.992&0.0009&1.9982&0.0003&1.9994&0&2\\
54&0.5991&123.0332&0.4806&512.1205&0.3817&1928.2614&0.2934&6796.8576\\
56&0.3266&110.1283&0.2159&345.4543&0.1452&1004.3413&0.1011&2798.7363\\
57&0.1427&66.3875&0.0595&138.6571&0.0368&269.1887&0.0166&522.0983\\
58&0.2261&91.8859&0.1241&227.1284&0.0685&522.7471&0.0352&1159.563\\
60&1&0&1&0&1&0&1&0\\
62&0.6806&115.2031&0.3775&511.3079&0.2104&1427.2255&0.1077&3330.3698\\
72&0.0059&6.6752&0.0014&7.1114&0.0004&7.5419&0.0001&8.7902\\
73&0.4241&153.995&0.2697&539.3759&0.1559&1611.5&0.0902&4281.3997\\
74&0.6732&112.9072&0.5043&513.9198&0.3615&1857.4982&0.2324&6055.0327\\
76&0.0027&1.9973&0.0008&2.1884&0.0003&2.3443&0&2.4521\\
77&0.0076&7.4463&0.0024&8.0492&0.0006&8.4614&0.0002&9.1361\\
78&0.0212&17.3528&0.0051&20.4336&0.002&23.2771&0.0005&24.8262\\
90&1&0&1&0&1&0&1&0\\
94&0.2209&119.264&0.1172&304.6733&0.0527&682.9739&0.0266&1295.4297\\
104&0.0594&47.047&0.016&62.2863&0.0051&73.0639&0.0012&74.4246\\
105&1&0&1&0&1&0&1&0\\
106&1&0&1&0&1&0&1&0\\
108&0.0133&11.6458&0.0033&12.9831&0.0014&14.3384&0.0001&15.3613\\
110&0.7892&85.207&0.7088&422.3375&0.6229&1918.6175&0.5404&8131.7471\\
122&0.7727&90.6083&0.6454&467.5212&0.515&2016.0089&0.3819&7780.4991\\
126&0.3602&115.4745&0.2572&389.2605&0.1823&1232.7578&0.1309&3752.0127\\
128&0.0042&4.767&0.0014&5.1223&0.0004&5.3654&0.0001&5.8271\\
130&0.0865&41.9356&0.0468&89.3382&0.0231&190.9837&0.0128&407.73\\
132&0.0086&7.517&0.0021&8.0448&0.0008&8.9295&0.0002&9.4588\\
134&0.4792&127.2468&0.3204&448.0948&0.2053&1372.3773&0.1371&3794.8008\\
136&0.0081&6.9283&0.0023&7.6935&0.001&9.3916&0.0001&8.8709\\
138&0.3932&121.1175&0.3025&429.4487&0.232&1456.7279&0.1803&4788.5434\\
140&0.0065&5.4538&0.0014&6.5541&0.0004&7.4921&0.0002&8.3589\\
142&0.6478&115.6484&0.4983&511.3795&0.376&1925.1316&0.2793&6605.3952\\
146&0.4318&126.3201&0.3181&451.5111&0.2391&1490.3106&0.1689&4798.1028\\
150&1&0&1&0&1&0&1&0\\
152&0.1055&57.6982&0.0559&126.3235&0.0308&271.1771&0.0155&580.6825\\
154&1&0&1&0&1&0&1&0\\
156&0.02&18.6955&0.0066&23.3457&0.0024&27.5274&0.0004&31.1357\\
160&0.019&14.9671&0.005&16.1459&0.0014&17.7011&0.0003&20.3167\\
162&0.1731&76.9699&0.1151&210.7525&0.0764&568.7733&0.0534&1517.3141\\
164&0.1176&64.8285&0.0527&130.9186&0.0232&228.8644&0.0075&371.3313\\
168&0.0949&67.665&0.0405&135.6538&0.0168&258.6673&0.0085&440.1266\\
170&1&0&1&0&1&0&1&0\\
172&0.0485&37.6656&0.0175&63.234&0.0068&99.9733&0.0026&151.9821\\
178&0.0135&11.9561&0.0034&13.2167&0.0016&14.1006&0.0001&13.7332\\
184&0.5655&126.3228&0.4373&500.8263&0.334&1775.3973&0.2422&5955.3257\\
200&0.0031&2.3074&0.0003&2.4128&0.0001&2.5936&0.0001&2.9415\\
204&0.002&0&0.0005&0&0.0004&0&0&0\\
232&0.0079&7.309&0.0024&8.2605&0.0007&8.5996&0.0001&8.5114\\
\end{longtable}

\bibliography{reflast}

\end{document}